\documentclass[%
aps,%
12pt,%
final,%
notitlepage,%
oneside,%
onecolumn,%
nobibnotes,%
nofootinbib,%
superscriptaddress,%
noshowpacs,%
centertags]%
{revtex4}
\usepackage[koi8-r]{inputenc}
\usepackage[english,russian]{babel}
\usepackage[T2A,T1]{fontenc}
\usepackage{epsfig}
\usepackage[centerlast,normalsize]{caption2}
\begin{document} 

~\hfill {\it Bulletin of the Lebedev Physics Institute, No. 12, 2006}
\rule{\textwidth}{0.1mm}

\vspace{5mm}
\begin{center}
{\large\bf ON A POSSIBLE MECHANISM}\\
{\large\bf OF THE PRODUCTION OF PENTAQUARK STATES 
$\Theta$(ud${\rm \bar s}$ud)}
\\ \vspace{3mm}
\label{st1}
S. P. Baranov\\
{\sl P.N.Lebedev Institute of Physics, Moscow 119991}
\\ \vspace{5mm}

\begin{minipage}{135mm}
{\bf
A new mechanism of the production of hypothetical
$\Theta({\bf ud{\bar s}ud})^+$ pentaquark state is suggested. 
The considered mechanism is based on the exchange by exotic Regge 
trajectories in the ${\bf t}$ channel. It is shown that the momentum 
distributions of $\Theta^+$ baryons can differ significantly from the 
ones of the ordinary hyperons. 
} \end{minipage} \end{center}
\rule{\textwidth}{0.1mm}

\large
Data in favor of the discovery of pentaquark states, and, in particular,
states of the $(ud{\bar s}ud)$ quark composition [1] are yet disputable 
and not generally acknowledged. The question on the character of 
kinematic distributions for these exotic baryons remains of the urgent
interest nowadays. Comparing results of the search for pentaquark states,
which are obtained in different experiments, we should allow for the fact
that the expected signal can correspond to different acceptance 
conditions of the experimental arrangements.

The expected characteristics of the pentaquark production are usually
considered to be similar to those of the $\Lambda$(1520) hyperon because 
they both are baryons and carry the strangeness. This analogy is true
for the kaon-exchange mechanism analyzed in [2]. However, this mechanism
is, generally speaking, may not be the only mechanism contributing to the 
pentaquark production, although it is necessary by virtue of the existence 
of the $\Theta pK$ interaction vertex responsible for the pentaquark decay. 
In the given case, this mechanism can even be not the dominant one.

In order to assemble the $\Theta^+$ pentaquark from a proton and a kaon, 
a rather complicated rearrangement of color fluxes is required. This 
complication, in a certain extent, explains the low probability of the 
inverse process, i.e., the small width of the decay
$\Theta^+ \to p {\bar K}^0$. On the other hand,
if we reject the condition of the indispensable use of all quarks from 
the incoming proton (and restrict our consideration only to one quark 
or diquark), then the color structure of the interaction is essentially
simplified. However, in this case, we have to additionally pick up the
missing quarks from the $t$ channel. Hence, the interaction must be
mediated by not a single kaon but a multiquark complex.

In the framework of the initial assumption that the exotic states do
exist, the hypothesis that interactions also can be mediated by exotic
hadrons seems to be quite natural. Assuming the parameters of the
corresponding Regge trajectories to be known, we can calculate both 
the energy dependence of the interaction cross section and the kinematic 
distributions of the produced particles. The absolute normalization of 
the cross section remains uncertain, since the interaction constants of 
exotic hadrons are unknown.%
\footnote{The $\Theta pK$ interaction vertex in the kaon exchange 
mechanism is considered to be known since it can be extracted from 
the $\Theta^+ \to p{\bar K}^0$ decay width.} 
In order to find the relevant parameters of exotic trajectories, one can
employ the hypothesis of the intercept additivity with respect to the 
quark composition. According to this hypothesis, every addition or 
subtraction of a quark shifts the trajectory by the same value. 
The application of this hypothesis is illustrated in Fig. 1. 
By adding and subtracting two quark-antiquark pairs, the pentaquark state 
is reduced to a set of two nucleons, one hyperon, and two mesons. Hence 
follows the relationship between the intercept of the exotic trajectory 
and the intercepts of the known non-exotic trajectories:
$\alpha_\Theta = 2\alpha_N + \alpha_\Lambda - 2\alpha_R$.
Thus, we manage to establish the functional form of the fragmentation 
functions for quarks and diquarks.

\begin{figure}
\epsfig{figure=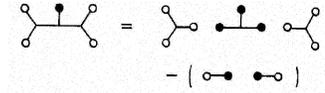, width=8.cm}
\caption{
Graphical method illustrating the derivation of the relationship 
$\alpha_\Theta = 2\alpha_N + \alpha_\Lambda - 2\alpha_R$.
} \end{figure}

The calculations were carried out within the formulism of the 
Quark--Gluon String model [3 -- 10]. For the sake of convenience, we 
here present the short list of the relevant formulas without repeating 
their derivations. The inclusive differential production cross section 
for a hadron of the type $h$ can be written in the form
\begin{equation}
\frac{d\sigma^h}{dp_{Tx} dp_{Tz}\,dy}=\sigma_{DD} \varphi_{DD}^h +
 \sum_{n=1}^{\infty}\,\sigma_n\varphi_n^h \label{sigma H},
\end{equation}
\noindent where $p_{Tx}$ and $p_{Tz}$ are the transverse momentum 
components, $y$ is the rapidity, and $\sigma_{DD}$ is the contribution 
of the diffraction dissociation (not taken into account in the present 
study). The summation over $n$ corresponds to the contributions of the 
$n$-Pomerons configurations in the exchange channel. The corresponding 
cross sections $\sigma_n$ have been calculated in [4]. The functions 
$\varphi_n$ describe the fragmentation of color strings into hadrons:
\begin{eqnarray}
\varphi_n^h&=&
a^h\,\bigl[ F_{val}^h(x_{+},n)F_{\overline{val}}^h(x_{-},n) \label{phi}
           +F_{\overline{val}}^h(x_{+},n)F_{val}^h(x_{-},n) + \nonumber \\
&&   +(n-1)[F_{sea}^h(x_{+},n)F_{\overline{sea}}^h(x_{-},n)
           +F_{\overline{sea}}^h(x_{+},n)F_{sea}^h(x_{-},n)]\bigr].
\end{eqnarray}
\noindent 
   Here, $x_{\pm}=\frac{1}{2}[(x_{\perp}^2+x_F^2)^{1/2} \pm x_F],\quad
       x_{\perp}=2m^h_{\perp}/\sqrt{s},$ \quad
      $x_{+}x_{-}=(m^h_{\perp})^2/s$, \quad $x_{+}-x_{-}=x_F.$\\
Each term in expression (2) corresponds to an individual color string
stretched between valence or sea partons of the interacting hadrons. The
contribution of each of the strings is represented in the form of the
product of two independent functions $F_i(x_{+})$ and $F_i(x_{-})$ 
corresponding to end partons of the given string. In turn, the functions  
$F(x_{\pm})$ are calculated as the convolution of parton distribution 
function $f_i(x,n)$, parton fragmentation function 
$D_i^h(x_{\pm}/x',\,p_T)$, and the weight function 
$T(x_F/x',\,p_T,\,n)$ modelling the transverse momentum:
\begin{equation}
F^h(x_{\pm},n)=\sum_i
\int_{x_{\pm}}^1
f_i(x',n)\,D_i^h(x_{\pm}/x',\,p_T)\,T(x_F/x',\,p_T,\,n)\,dx'.
\end{equation}
\noindent As in the usual parton model, the distribution function 
determines the probability to find in the original hadron a parton 
of the type $i$ carrying the longitudinal momentum fraction $x$, while 
the fragmentation function determines the probability of transforming 
the given parton into the hadron $h$ of the final state.

In the case of the $\Theta(ud\bar{s}ud)^+$-pentaquark production which 
we are interested in, the calculations were performed as for a usual 
baryon, e.g., the charmed $\Lambda_c^+$ hyperon where the charmed quark 
is substituted with the $(\bar{s}ud)$ triquark complex. In this case, 
in accordance with the above considerations related to the intercepts of 
exotic trajectories, the quark and diquark fragmentation functions take
the form
\begin{eqnarray}
D_{q\to\Theta+X}(z,p_T) &=& c_1\,(1-z)^{\alpha_R-2\alpha_\Theta+\lambda}, \\
D_{qq\to\Theta+X}(z,p_T) &=& c_2\,(1-z)^{-\alpha_Q+\lambda}.
\end{eqnarray}
\noindent Here, $\alpha_\Theta = 2\alpha_N + \alpha_\Lambda - 2\alpha_R$
and $\alpha_Q = 2(\alpha_N + \alpha_\Lambda) - 3\alpha_R$,\quad where 
$\alpha_N$, $\alpha_\Lambda$, and $\alpha_R$ are the intercepts of the 
nucleon, hyperon and $\rho$-meson trajectories, 
$\lambda=2\alpha'_Rp_T^2$, and $\alpha'_R=1$ GeV$^{-2}$ is the slope
of the Reggeon trajectory. The normalization coefficients $c_1$ and $c_2$ 
are not determined in the theory and remain to be free parameters. The 
full list of parton distribution functions and parton fragmentation 
functions for non-exotic states can be found in [9, 10]. The calculations
were performed using the Fortran code HIPPOPO described in detail in [11].

\begin{figure}
\epsfig{figure=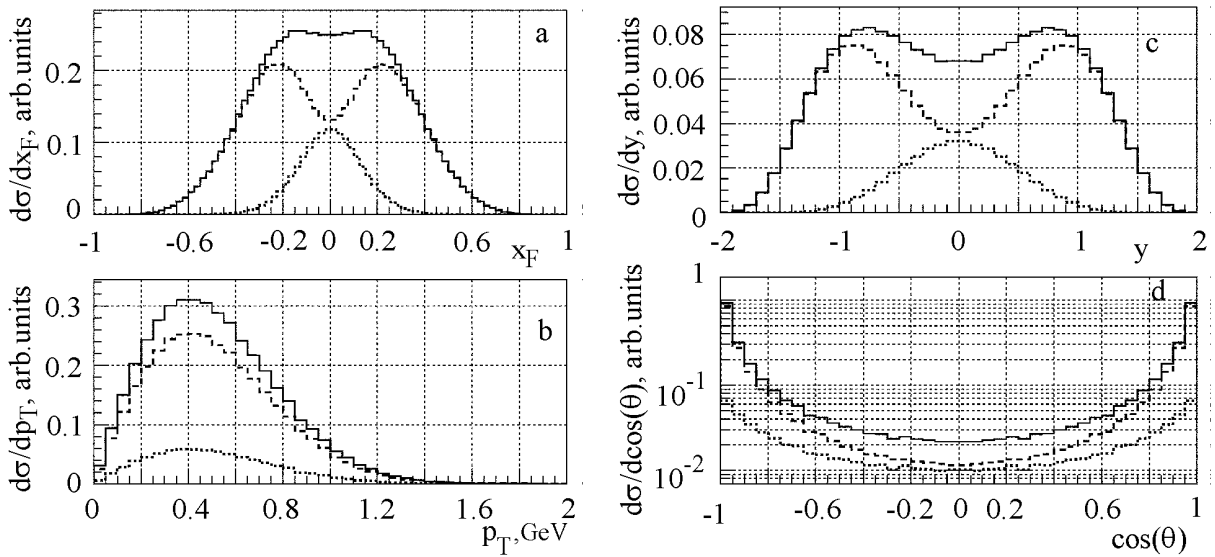, width=18.5cm}
\caption{
Kinematic distributions of $\Theta^+$ pentaquarks produced in $pp$
collisions at $\sqrt{s} = 12$ GeV over (a) the Feynman variable $x_F$;
(b) the transverse momentum $p_T$; (c) the rapidity $y$; and (d) the
cosine cos$(\theta$) of the pentaquark emission angle $(\theta)$ in 
the $pp$ center-of-mass system. Dotted and dashed curves correspond to
the contribution of the quark and diquark fragmentation, respectively;
solid curves represent the sum of all contributions.
} \end{figure}

In Fig. 2, predictions are shown for the $\Theta^+$-pentaquark momentum 
distributions as functions of the Feynman variable $x_F$, transverse 
momentum $p_T$, rapidity $y$, and the cosine $\cos(\theta)$ of the 
emission angle $\theta$ in the colliding protons center-of-mass system 
at the total initial energy $\sqrt{s} = 12$ GeV. The relation between 
the coefficients $c_1$ and $c_2$ responsible for the quark and diquark 
fragmentation contributions was taken to be $c_1$:$c_2$ = 1:10, i.e., 
approximately the same as it had been obtained from the data related 
to the production of ordinary baryons. The exchange by a heavier Regge 
trajectory results in an essential shift of the $\Theta^+$-pentaquark 
spectrum compared to both the $\Lambda$ hyperon spectrum and the 
predictions of the kaon exchange model [2]. In the latter case, the 
maximum of the momentum distribution lies in the region $x_F\simeq 0.85$. 
With the increase in the beam energy, the spectrum displaces to the 
region of lower $x_F$ values. This is the consequence of just the 
kinematics (translating the Lorentz-invariant cross section 
$d\sigma/dy$ into the noninvariant cross section $d\sigma/dx_F$).

Thus, we can see that using the hypothesis on the exchange by exotic 
heavy trajectories leads to a significant variation in the shape of 
momentum distribution for the state being considered. This fact should 
be taken unto account in the analysis of experimental data. 
Whether our prediction correspond to the truth, is the question 
that should be solved in the future together with the confirmation 
of the very existence of pentaquark baryons.

The author is grateful to P. F. Ermolov for initiating interest in the
study.

\end{document}